\documentclass[AMA,STIX1COL]{WileyNJD-v2}

\newcommand\BibTeX{{\rmfamily B\kern-.05em \textsc{i\kern-.025em b}\kern-.08em
T\kern-.1667em\lower.7ex\hbox{E}\kern-.125emX}}

\articletype{Research Article}

\received{<day> <Month>, <year>}
\revised{<day> <Month>, <year>}
\accepted{<day> <Month>, <year>}

\begin{document}

\title{\text{L}earning Optimal Biomarker-Guided Treatment Policy for Chronic Disorders}

\author[1]{Bin Yang}

\author[1]{Xingche Guo}

\author[2]{Ji Meng Loh}

\author[1]{Qinxia Wang}

\author[1,3]{Yuanjia Wang}

\authormark{Yang \textsc{et al}}

\address[1]{\orgdiv{Department of Biostatistics}, \orgname{ Columbia University}, \orgaddress{\state{New York}, \country{U.S.A}}}

\address[2]{\orgdiv{Department of Mathematical Sciences }, \orgname{New Jersey Institute of Technology}, \orgaddress{\state{New Jersey}, \country{U.S.A}}}

\address[3]{\orgdiv{Department of Psychiatry}, \orgname{Columbia University}, \orgaddress{\state{New York}, \country{U.S.A}}}

\corres{Yuanjia Wang, 722 W168th Street, New York, NY 10032. \email{yw2016@cumc.columbia.edu}}


\abstract[Abstract]{Electroencephalogram (EEG) provides noninvasive measures of brain activity and is found to be valuable for diagnosis of some chronic disorders. Specifically, pre-treatment EEG signals in alpha and theta frequency bands have demonstrated some association with anti-depressant response, which is well-known to have low response rate. We aim to design an integrated pipeline that improves the response rate of major depressive disorder patients by developing an individualized treatment policy guided by the resting state pre-treatment EEG recordings and other treatment effects modifiers. We first design an innovative automatic site-specific EEG preprocessing pipeline to extract features that possess stronger signals compared with raw data. We then estimate the conditional average treatment effect using causal forests, and use a doubly robust technique to improve the efficiency in the estimation of the average treatment effect. We present evidence of heterogeneity in the treatment effect and the modifying power of EEG features as well as a significant average treatment effect, a result that cannot be obtained by conventional methods. Finally, we employ an efficient policy learning algorithm to learn an optimal depth-2 treatment assignment decision tree and compare its performance with Q-Learning and outcome-weighted learning via simulation studies and an application to a large multi-site, double-blind randomized controlled clinical trial, EMBARC.}

\keywords{Causal forests; Heterogeneous treatment effect; Treatment policy; Electroencephalography; Biomarkers; Major Depressive Disorder}

\jnlcitation{\cname{%
\author{Yang B}, 
\author{Guo X}, 
\author{Loh JM}, 
\author{Wang Q}, and 
\author{Wang Y} (\cyear{2023}), 
\ctitle{Learning Optimal Biomarker-Guided Treatment Policy for Chronic Disorders}, \cjournal{Statistics in Medicine}, \cvol{2017;00:1--6}.}}

\maketitle

\section{Introduction}\label{sec1}
Major depressive disorder (MDD) is a chronic or episodic mood disorder that accounts for substantial disability and mortality, yet the antidepressant treatment response is unpredictable and varies across depressive patients \citep{fava2003background}. With technological advancements in collecting biomarkers, it is desirable to systematically explore a large variety of biosignatures to examine the heterogeneity of treatment effect (HTE) and to use biomarkers (e.g. electroencephalography) to guide individualized treatment selection. EEG provides tools to study the temporal dynamics of the brain and its cognitive processes with the advantage of being non-invasive, economic, and having high temporal resolution. Quantitative EEG measures have been shown to distinguish patients affected by  neurological and psychiatric disorders including dementia and depression from health controls \citep{hughes1999conventional}. In addition, previous research has demonstrated the association between signals from resting-state electroencephalography (rsEEG) and antidepressant response \citep{bruder2013electrophysiological,wu2020electroencephalographic}. More specifically,  Bruder et al. (2013)\cite{bruder2013electrophysiological} demonstrated associations between pre-treatmernt EEG measures in alpha and theta bands and antidepressant response. Korb et al. (2009) \cite{korb2009rostral} discovered evidence that higher pre-treatment resting state theta EEG activity in the rostral ACC and antidepressant response are significantly correlated.

A multi-site, placebo-controlled randomized clinical trial, \textbf{E}stablishing \textbf{M}oderators and \textbf{B}iosignatures of \textbf{A}ntidepressant \textbf{R}esponse in \textbf{C}linical care 
$\left( \text{EMBARC} \right)$, was designed to systematically explore the moderating effects of biomarkers on the outcome of antidepressant treatment, where a large variety of clinical, biological (e.g., neuroimaging and electrophysiology measures) and psychosocial markers were collected \citep{trivedi2016establishing}.The extensive biomakers collection from EMBARC provides a unique opportunity to facilitate estimation of treatment effects and treatment assignment policy for MDD patients. 

 One challenge of incorporating EEG biomarkers into treatment decision-making is that due to site differences in acquisition procedures and equipment, it is difficult to employ a unified EEG preprocessing and feature extraction pipeline that yields standardized output \citep{bigdely2015prep}. Increasing attention has been devoted to EEG data preprocessing algorithms that can automatically perform artifact detection and repair. Bigdely-Shamlo et al. (2015) \cite{bigdely2015prep} demonstrated the importance of early-stage EEG preprocessing and proposed a standardized early-stage EEG processing pipeline (PREP) by adopting a non-filtering line-noise removal, robust re-referencing bad channel removal and interpolation procedure. Jas et al. (2016) \cite{jas2016automated} developed algorithms for automatic bad trials rejection and repair by finding the optimal rejection threshold in a data-driven way. However, none of the existing work evaluates the impact of preprocessing on detecting HTE. 


In this work, we first design a site-specific EEG preprocessing pipeline that accommodates equipment differences. We tackle the noisy channel issue by employing methods from Bigdely-Shamlo et al. (2015) \cite{bigdely2015prep} for bad channel detection and interpolation. We further adopt the algorithms developed by Jas et al. (2016) \cite{jas2016automated} and incorporate an automatic bad channel and bad epoch identification process into our pipeline. After obtaining processed EEG data, the relative theta and alpha band power from the EEG recordings are then extracted using the multitaper method. This automatic pipeline reduces errors caused by manual inspection and ensures that the extracted features are standardized and reproducible. We compare the extracted features from the processed and raw data, and show that patterns in the processed data are more evident. We also contrast the average signal before and after preprocessing 
and validate the utility of pre-processing  by comparing the HTE estimated from raw and processed features. 

Second, it is unknown what is the best approach to estimate HTE and learn an optimal treatment policy from the available EEG features, which are correlated high-dimensional biomarkers with potential nonlinear interactions. Instead of using a region-based approach as in the previous work \citep{bruder2013electrophysiological,wu2020electroencephalographic}, we aim to simultaneously analyze the relative theta and alpha band power from all brain regions. There are emerging statistical and machine learning literatures that focuses on HTE estimation with complex form, with various techniques including LASSO \citep{tian2014simple}, BART \citep{logan2019decision,hill2011bayesian}, and neural networks \citep{shi2019adapting,farrell2021deep}. Causal forests, extending the Breiman's random forest algorithm, provides a powerful and non-parametric consistent estimation for the CATE \citep{wager2018estimation} and are known to performs well when there exists many possibly nonlinear interactions between covariates and has valid asymptotic properties and inference \citep{athey2019generalized}. In this work, we use causal forests to explore and estimate the heterogeneous treatment effects. We also leverage the extracted EEG features as well as the clinical, demographic markers to compute a doubly robust estimate of the average treatment effect (ATE) of sertraline, and explore whether the EEG bio-signatures modify the treatment effects.   

Finally, after identifying the heterogeneity of the treatment effect, it is desirable to learn a simple treatment policy that assigns the optimal treatment to maximize the expected utility guided by biomarkers, which serves as the ultimate goal of this work as illustrated in \hyperref[figure A1]{Figure A1}. Many algorithms have been developed to learn a treatment assignment policy. Nahum-Shani et al. (2012) \cite{nahum2012q} introduced Q-learning under a regression model to estimate the optimal sequence of decision rules. Zhao et al. (2012) \cite{zhao2012estimating} proposed outcome-weighted learning that directly estimates treatment rule without requiring modeling outcomes. Liu et al. (2018) \cite{liu2018augmented} combined Q-learning and outcome-weighted learning under a double robustness framework. These methods provide a nonparametric treatment policy which may be difficult to implement in clinics. Our goal is to learn an interpretable tree-based policy which is easy for clinicians to implement. We achieve this goal by employing an efficient policy learning algorithm by Athey and Wager (2021) \citep{athey2021policy} that has established strong guarantees for the asymptotic utilitarian regret of the resulting policy. 
To demonstrate the superiority of the policy learning algorithm, we compare its performance with policies learned via Q-learning and outcome-weighted learning in synthetic simulation studies mimicking real data and applications to EMBARC \citep{trivedi2016establishing}. 
 
The remainder of this article is organized as follows. In \hyperref[Section 2]{Section 2}, we first present an overview of the causal parameters of interest and formulate the treatment effects estimation and policy learning algorithms. We then illustrate our site-specific automatic EEG preprcoessing and feature extraction pipeline. In \hyperref[Section 3]{Section 3} we perform simulations based on real data and demonstrate the superiority of our implemented policy learning compared to Q-learning and outcome-weighted learning. In \hyperref[Section 4]{Section 4}, through an application of our method to a real world study - EMBARC \citep{trivedi2016establishing}, we present evidence of heterogeneity in the treatment effect and the modifying power of EEG features as well as a significant average treatment effect. We also learn an optimal depth-2 treatment assignment decision tree where EEG biomarkers are selected as splitting variable. Lastly, in \hyperref[Section 5]{Section 5} we present a discussion of our integrated pipeline and its utility for treatment assignment decision-making.

\section{Methods}
\label{Section 2}

\subsection{Causal Forests}

Consider $n$ independent and identically distributed subjects. For each subject, let $X_i$ denote a pre-treatment feature vector and $W_i \in \{0, 1\}$ denote a treatment indicator where $W_i = 0$ indicates  the control treatment and $W_i = 1$ the active treatment. Causal effects are defined via the potential outcomes as in Imbens and Rubin (2015) \cite{imbens2015causal}. Let $Y_i(0), Y_i(1)$ denote the potential outcomes we would have observed had the subject been assigned to control or treatment, respectively. The casual effect at the subject level is: $\tau_i = Y_i(1) - Y_i(0)$. The conditional average treatment effect (CATE) is defined to be $\tau(x) = \mathbb{E}[Y_i(1) - Y_i(0)|X_i = x]$. 


Casual Forests is built upon the random forest algorithms introduced by Breiman (2001) \cite{breiman2001random}, an ensemble model that makes prediction by averaging predictions from individual trees and allows for high dimensional interactions. It provides a powerful non-parametric consistent estimation for the heterogeneous treatment effects. Under unconfoundedness assumptions, i.e., $\{Y_i^{(1)},Y_i^{(0)}\} \perp \!\!\! \perp W_i | X_i$, Causal Forests seeks to maximize the difference in treatment effect between the two child nodes, while controlling for optimism and keeping the forests ``honest" \citep{wager2018estimation}. Specifically, let $e(x) = \mathbf{P}[W_i|X_i = x]$ be the propensity score and $m(x) = \mathbf{E}[Y_i|X_i =x]$ be the expected outcome given covariates marginalized over treatments. We use the Causal Forests implemented by the R-learner, which consists of the following steps: \\
1) Estimate $\hat{m}(\cdot)$,  $\hat{e}(\cdot)$ by fitting two separate regression forests. \\
2) Grow a forest via 
$$\hat{\tau}(x) = \frac{\sum^n_{i=1}\alpha_i(x)(Y_i-\hat{m}^{-i}(X_i))(W_i -\hat{e}^{-i}(X_i))}
{\sum^n_{i=1}\alpha_i(x)(W_i -\hat{e}^{-i}(X_i))^2},$$
where $\alpha_i(x) = \frac{1}{B} \sum^B_{b=1}\frac{\text{I}({X_i\in L_b(x),i\in B}}{|{i:X_i\in L_b(x),i\in B}|}$ is the learned adaptive weights where we denote $B$ to be the total number of trees and $L_b(x)$ to be the terminal leaf. The tuning parameters are automatically selected via cross-validation by minimizing an ``R-learner" objective function \citep{nie2021quasi}, 
$$\hat{\tau}(\cdot) = \text{argmin}_\tau\left\{\sum^{n}_{i=1}((Y_i-\hat{m}^{-i}(X_i)) - \tau(X_i)(W_i -\hat{e}^{-i}(X_i))^2 + \Lambda_n(\tau(\cdot)) \right\},$$
where $\Lambda_n(\tau(\cdot))$ is a regularizer that controls the complexity of learned $\hat\tau(\cdot)$ function \citep{nie2021quasi}.  

\subsection{Estimation of the ATE}

An efficient doubly robust estimation of ATE can be obtained by plugging in the causal forests predictions. As discussed in Chernozhukov et al. (2018a) \cite{chernozhukov2018double}, under regularity conditions, such approaches can yield semiparametrically efficient ATE estimates and accurate standard error estimates. When either the propensity score model or the outcome model is correctly specified, the estimator is consistent. If both models are correctly specified, the efficiency is improved. In a randomized trial, propensity scores are available or can be consistently estimated. Thus, it is highly desirable to use a doublyrobust method to estimate ATE to improve efficiency.  The exact expression for the average treatment effect $\hat\tau$ is: 
$$\hat\tau = \frac{1}{n}\sum^{n}_{i=1} \hat{\Gamma}_i,$$ where
$$\hat{\Gamma}_i = \hat{\tau}^{(-i)}(X_i) + \frac{W_i-\hat{e}^{(-i)}(X_i)}{\hat{e}^{(-i)}(X_i)[1-\hat{e}^{(-i)}(X_i)]} [Y_i - \hat{m}^{(-i)}(X_i) - [W_i - \hat{e}^{(-i)}(X_i)]\hat{\tau}^{(-i)}]\citep{athey2019estimating}.
$$
Here $\hat{\tau}^{(-i)}(X_i)$ is the treatment effect estimate,
$\hat{e}^{(-i)}(X_i) = \mathbb{E}[W_i|X_i = x]$ is the propensity score for subject $i$ estimated using a separate regression tree, and $\hat{u}^{(-i)}(X_i,W_i) = \hat{m}^{(-i)}(X_i) + [W_i - \hat{e}^{(-i)}(X_i)]\hat{\tau}^{(-i)}$ is an estimate of the conditional mean for subject $i$, where $m(X_i) = \mathbb{E}[Y|X_i]$ is also estimated using a separate regression tree. 

\subsection{Learning Treatment Policies}

When treatment effect heterogeneity is detected, it is desirable to learn an interpretable treatment policy depending on the observed characteristics so that a clinician can assign the optimal treatment to individuals and benefit the target population. Formally, we would like to learn a policy $\pi$ that maps observed subjects' characteristics $X_i \in \mathcal{X}$ to a binary treatment $W_i$: $\pi: \mathcal{X} \rightarrow \{0, 1\},$ where $0$ indicates control treatment and $1$ indicates active treatment. 

Athey and Wager (2021) \cite{athey2021policy} developed a family of algorithms that are based on semiparametrically efficient estimation \citep{chernozhukov2022locally} (i.e., policy learning), which shows that one can construct efficient estimates of average-treatment-effect-like parameter $\theta$ as:
$$\hat{\theta} = \frac{1}{n}\sum^{n}_{i=1} \hat{\Gamma}_i,$$
where $\hat{\Gamma}_i$ is the doubly robust score appropriate for the target estimand. Athey and Wager (2021) \cite{athey2021policy} shows that  an optimal policy $\hat{\pi}$ given a pre-specified policy class $\Pi$(problem specific constraints, e.g., finite depth decision trees) solves:
$$\hat{\pi} = \text{argmax}\left\{\frac{1}{n}\sum^{n}_{i=1}(2\pi(X_i)-1)\hat{\Gamma}_i: \pi \in \Pi \right\},$$
with strong guarantees for the asymptotic utiliatarian regret of the resulting policy. We use the R package $\textbf{policytree}$ for the implementation of the above algorithms, where the learned policies belong to the class $\Pi$ of depth-$2$ decision tress \citep{sverdrup2020policytree}. 


Q-Learning is an alternative approach to estimate an optimal treatment policy. Q-learning utilizes regression analysis to assess the relative quality of intervention options and to estimate the optimal sequence of decision rule in settings where a sequential decision of intervention options are made \citep{nahum2012q}. In the case of a single stage decision learning, denote outcome of interest as $Y$, baseline covariates as $O_1$, and the intervention as $A_1$, the optimal decision rule is $d_1^*(O_1) = \text{arg max}_{a1}\mathbb{E}[Y|O_1, a_1]$. To implement Q-learning, we fit a Lasso regression by including all covariates and the interaction terms between treatment and covariates, and then estimate the optimal intervention as the one that leads to a higher outcome. 


In addition to Q-learning, we will compare policy learning with to outcome weighted learning (O-learning). In contrast to Q-learning, outcome weighted learning learns the optimal individualized treatment rule by directly maximizing the value function $\mathcal{V(D)} = E_{\mathcal{D}}[R]$, which is the expected outcome under the the treatment decision function. This is shown by Zhao et al. (2012)\cite{zhao2012estimating} to be equivalent to minimizing a weighted classification error. Denote $A \in \mathcal{A} = \{-1, 1\}$ to be the treatment assignments, $R$ to be the reward outcome, $H$ to be the patient's health history information, $D$ to be the treatment decision function that maps $H$ to the domain of $A$ and $\pi(A, H) = P(A=a|H = h)$ to be the treatment assignment probability. We adopt the single stage outcome weighted learning algorithm by Liu et al. (2018)\cite{liu2018augmented}, where the authors proposed that minimizing   
$$E\left[\frac{|R - s(H)|}{\pi(A, H)}I(A \text{ sign}(R-s(H)) \ne \mathcal{D}(H) \right]$$
reduces the variability of weights and improves the overall algorithm performance by subtracting an arbitrary function $s(H)$ of the feature variables. The goal is then to learn an optimal treatment decision function $f(\cdot)$ and set $D(H) = \text{sign}(f(H))$. To minimize the weighted classification error, surrogate loss functions can be used to replace the $0-1$ loss. We use the R package $\textbf{DTRlearn2}$ \citep{chen1dtrlearn2} for this task, where the binomial deviance loss is chosen. Note we do not use SVM with hinge loss, since the radial basis function (RBF) kernel did not perform well due to the simulation setup.

\subsection{EEG Preprocessing and Feature Extraction}


It is well known that the quality of pre-processing biomarkers including EEG could impact downstream analysis \citep{bigdely2015prep}. Here we examine how  different methods of preprocessing EEG may impact building a policy tree and estimating HTE. In EMBARC, resting-state EEG (rsEEG) were recorded from each of the four study sites, and they were recorded in four two-minute blocks, corresponding to an order of eyes open, eyes closed, eyes closed and eyes open. We design a site specific automatic EEG data preprocessing pipeline to minimize the biases. The general preprocessing steps are as followed: 1) The EEG data are resampled to 250 Hz. 2) Line noises at 60 Hz are removed. 3) High-pass filter at 1 Hz and low-pass filter at 50 Hz are applied. 4) We utilize four measures for automatic noisy channel detection: extreme amplitudes (deviation criterion), lack of correlation with any other channel (correlation criterion), lack of predictability by other channels (predictability criterion), and unusual high frequency noise (noisiness criterion) \citep{bigdely2015prep}. The bad channels are further interpolated. 5) Eye-blink related artifacts are removed using independent component analysis (ICA). 6) EEG data are segmented into two seconds epochs. 7) Bad epochs are automatically rejected with a sensor-specific threshold learned from cross validation \citep{jas2016automated}. 8) The filtered EEG data are re-referenced to the common average. 9) The 54 channels common to all sites are identified and extracted. 

\begin{figure}[h]
\label{figure 1}
\centering
\includegraphics[width=5in]{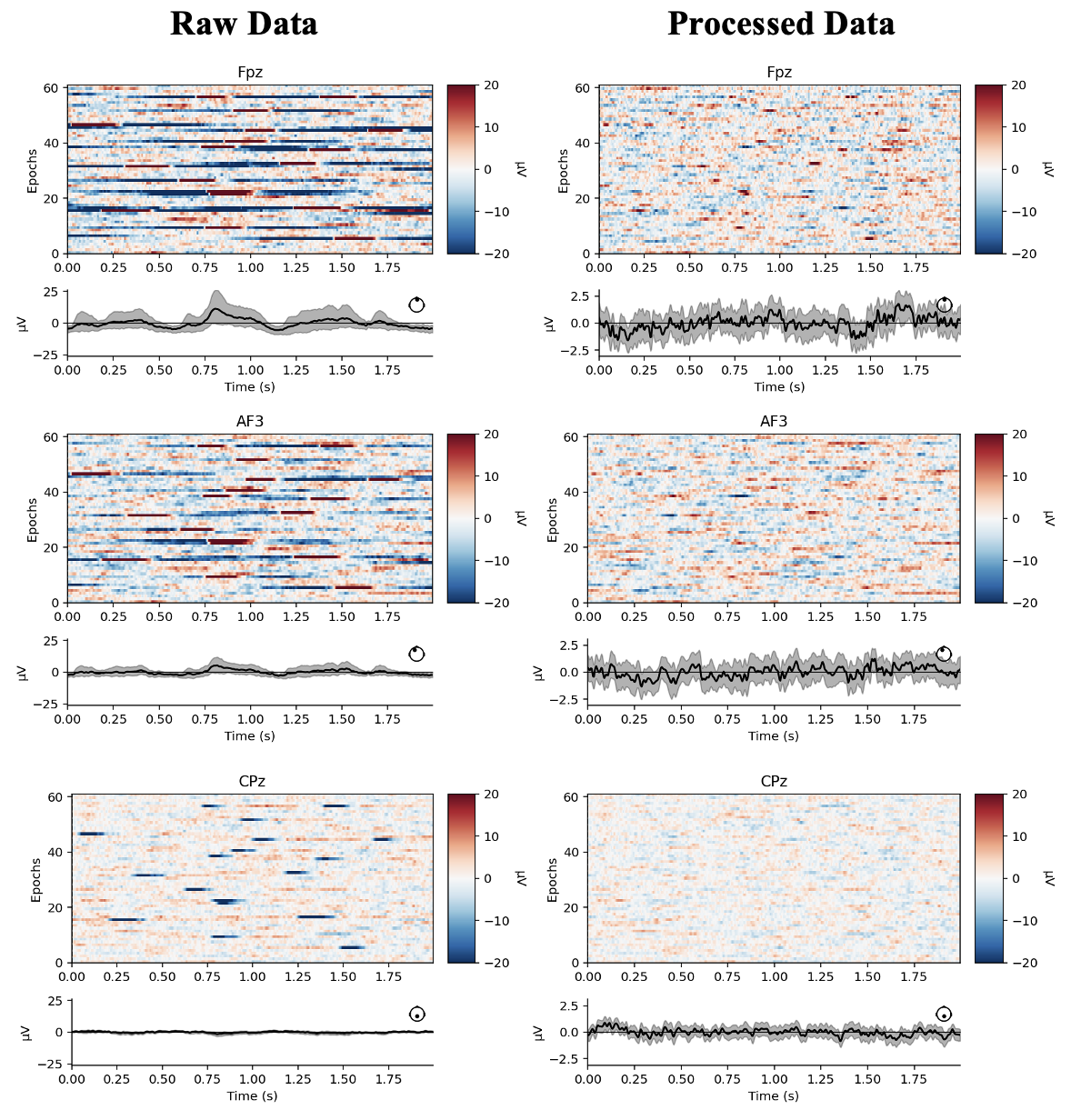}
\caption{Heat maps of EEG power of the same subject for three selected electrodes (Fpz, AF3 and  Cpz) with each row representing an epoch. Left column shows raw data and right column shows processed data. }
\end{figure}

Note that key steps such as 4 and 7 were not performed in existing literature on creating biomarker signatures of antidepressant response \citep{wu2020electroencephalographic}. To illustrate the value of our preprocessing pipeline, we use several visualization tools to contrast the raw and processed data. As shown in \hyperref[figure A2]{Figure A2}, we first compare the data in aggregate via the average signal over epochs. We notice that the average signal of processed data is less noisy. We then utilize heatmaps to contrast the correlation matrix for raw features and processed features in \hyperref[figure A3]{Figure A3}. We observe that the correlation between similar features increased, and the correlation decreased for dissimilar features. For instance, relative alpha band power during eyes open period in frontal and occipital regions are shown to have high correlation in the raw data set, and their correlation is further increased in the processed features. Moreover, relative alpha band power and relative theta band power in these regions demonstrate low to negative correlation in the raw data set, and their correlation further decreased in the processed features. After examining data in an aggregate way, we then focus on comparison at the channel level. As demonstrated in \hyperref[figure 1]{Figure 1}, we showcase the raw and processed EEG data for three selected electrodes (Fpz, AF3 and  Cpz).  An immediate observation shows that artifacts are removed so that the processed features are less noisy. Finally, we also demonstrate the improved ability of processed features for estimating the CATE in \hyperref[Section 4]{Section 4}, as illustrated by a smaller MSE on the transformed outcomes used to estimate the CATE. 

For feature extraction, we first decompose the EEG signals into functionally distinct frequency bands. We achieve this by utilizing the multitaper method which is a spectral analysis method that provides a robust spectral estimation. We then aim to extract features that could reflect frequency bands activity, and we are specifically interested in the theta (4-7Hz) and alpha (8-12Hz) frequency bands. To minimize the biases caused by  different EEG acquisition equipment at each study site, we extract the relative theta, alpha band power for each channel in the eyes open and eyes closed conditions, achieved through composite Simpson's rule. In total there are 216 EEG features for the alpha and theta band relative power of the common 54 channels in the eyes open and eyes closed conditions.

\section{Simulation Studies} 
\label{Section 3}

To demonstrate the superiority of our employed policy learning algorithms, we compare the aforementioned doubly robust scoring techniques with the treatment assignment policies that are learned via Q-learning implemented with LASSO regression and outcome weighted learning with binomial deviance loss in several synthetic data sets based on real data. The simulation settings are constructed based on the features from  EEG markers and clinical, demographic markers from the EMBARC study. We simulate the features in each  setting to mimic the real data set and simulate the treatment effect based on a tree structure with effect size estimated from the real data. Specifically, there are $254$ continuous covariates, including $216$ EEG features and $38$ demographic and clinical features and $10$ categorical demographic and clinical features.  The continuous covariates are simulated based on a multivariate normal distribution with mean and covariance matrix estimated from the real data set. The categorical covariates are simulated from the multinomial or Bernoulli distribution with class probability estimated from the real data set. We also simulate the treatment effect via a tree structure estimated from the real data: in the strong effect size setting, for each leaf node, we use the average of the estimated doubly robust scores $\hat\Gamma_{i,1}, \hat\Gamma_{i,0}$ as an estimator for $\mathbb{E}[Y(1)],\mathbb{E}[Y(0)]$ respectively. The treatment effect in the leaf node depends on the splitting variable and we simulate this relationship via a series of indicator function. In the weak effect size setting, we retain the same tree structure as the strong effect size setting but decrease the size of the treatment effect: in the leaf node where the optimal treatment is sertraline, we reduce $E[Y(1)]$ and increase $E[Y(0)]$ by 0.1, and do the opposite for the leaf node where the optimal treatment is placebo. 

\begin{figure}[ht]
\label{figure 2}
\centering
\includegraphics[width=6in]{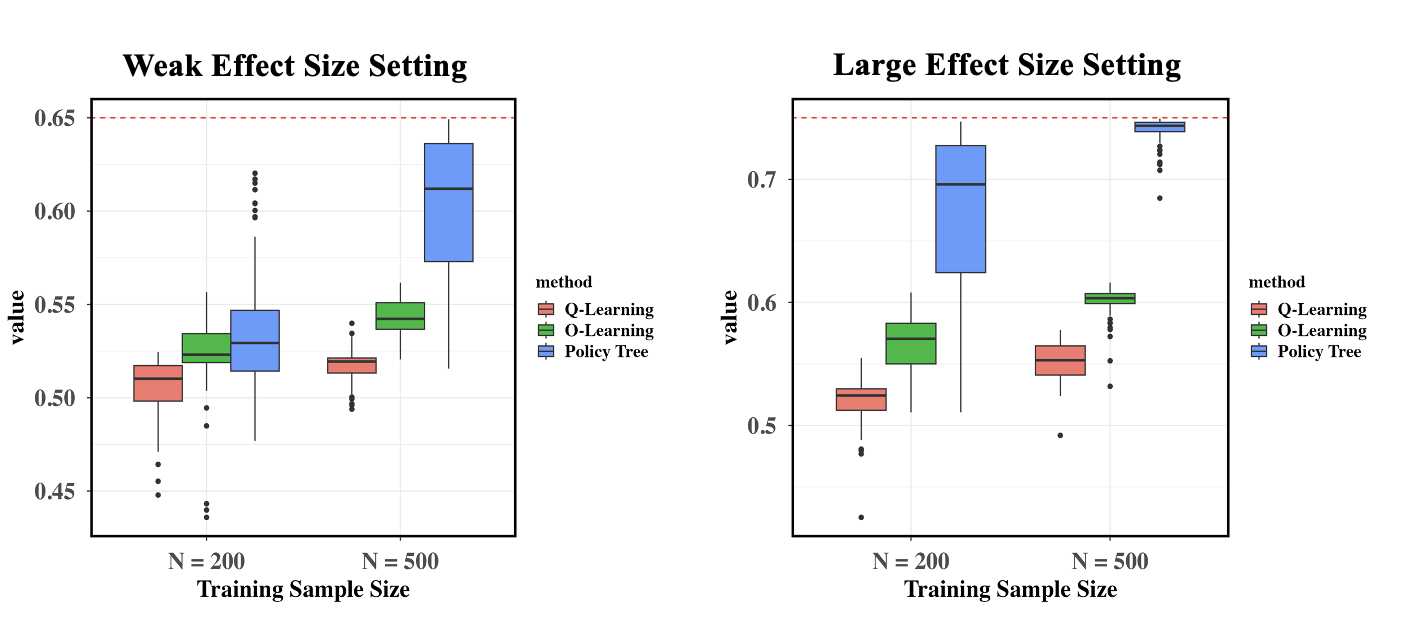}
\caption{Boxplot of value estimated in the testing set. Left-hand side plot summarizes results in the weak effect size setting and the right-hand side plot summarizes results in the large effect size setting. The red dashed lines denote the optimal value in the testing set.Red dashed line denotes the optimal value.}
\end{figure}

\begin{figure}[h]
\label{figure 3}
\centering
\includegraphics[width=6in]{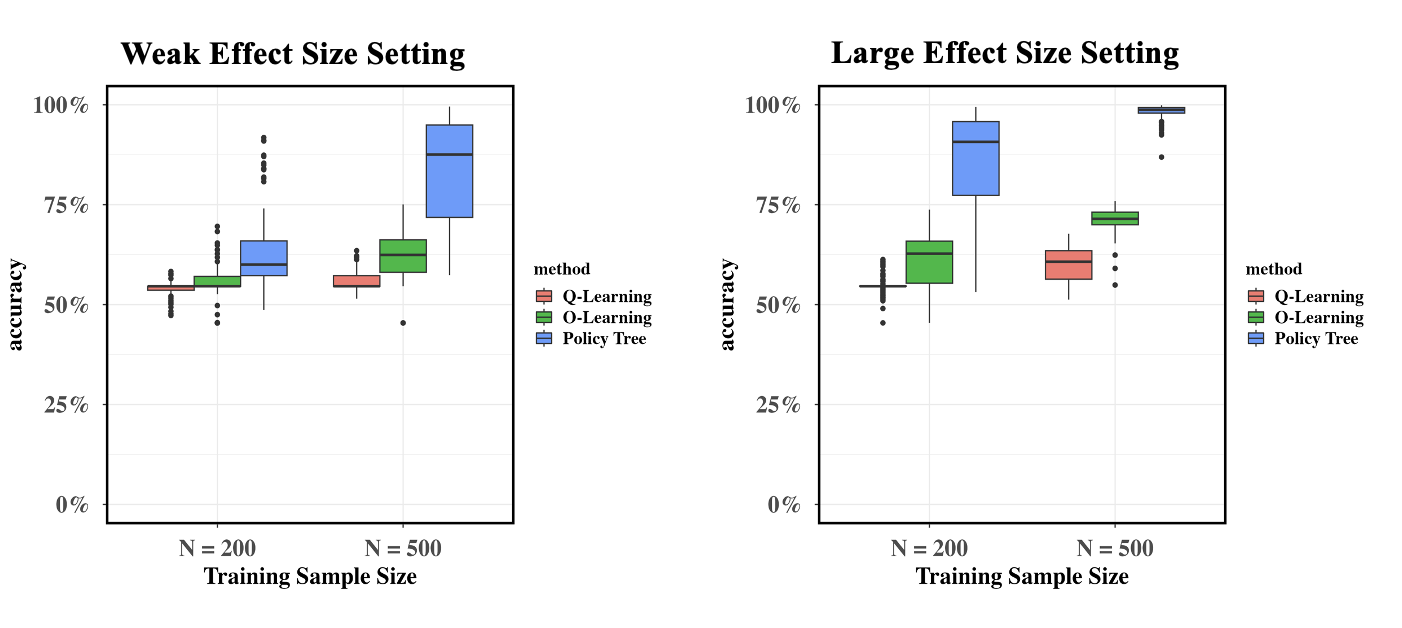}
\caption{Boxplot of classification accuracy in the testing set. Left-hand side plot summarizes results in the weak effect size setting and the right-hand side plot summarizes results in the large effect size setting.}
\end{figure}

The performance of the algorithms is evaluated via the value function, which is defined as: $V(\pi) = \mathbb{E}[Y_i(\pi(X_i))]$, as well as the accuracy of assigning subjects to their optimal treatment. For each effect size setting, we consider using training sample size $N = 200$ and $N = 500$. The algorithm performance is then evaluated in $100$ large, independent testing set with $50,000$ subjects. 
The results are summarized in the following \hyperref[figure 2]{Figure 2},  \hyperref[figure 3]{Figure 3}. In both the weak and large effect size settings, causal forests algorithm performs better than Q-learning and O-Learning in terms of value and treatment assignment accuracy in the testing set. Moreover, we see a substantial improvement in performance as the sample size in the training set increases.

\section{Application to EMBARC}
\label{Section 4}
EMBARC is a large multi-site, placebo-controlled randomized clinical trial. The 309 participants with MDD are randomized to selective reuptake inhibitor (SSRI), sertraline or placebo during the stage 1 of the trial \citep{trivedi2016establishing}. Participants are recruited at four study centers, University of Texas Southwestern Medical Center (Coordinating Center), Columbia University/Stony Brook (Data Center), Massachusetts General Hospital (MGH), University of Michigan, University of Pittsburgh, and McLean Hospital. Participants' clinical and biological markers are comprehensively assessed. For our analysis, 
we leverage the resting state, pre-treatment rsEEG recordings, as well as the clinical and demographic variables measured from EMBARC \citep{trivedi2016establishing} to examine the relative importance of rsEEG biomarkers on the HTE of the antidepressant. As illustrated in \hyperref[figure A1]{Figure A1} in the Appendix, the ultimate goal is to design an integrated pipeline from EEG preprocessing and feature extraction to constructing an optimal treatment assignment policy guided by useful EEG biomarkers with simplicity and interpretability.

Our primary clinical outcome of interest is a binary variable, the response at exit (HRSD score reduction of $50\%$ or greater). Clinical markers that demonstrated some predictive power for treatment outcome are included, such as atypical depression, anger attacks, Axis II disorders, etc. \citep{trivedi2016establishing}. 

\noindent $\underline{\textbf{Data Preprocessing and Model Training:}}$ We remove subjects with failure at baseline EEG quality control and missing outcome. We impute the missing clinical and demographic variables using the multiple imputation method. We then randomly split the data and use $70\%$ of the data as training set and the remaining $30\%$ of the data as the testing set. In addition, the minority class of response status at the exist in the training set is upsampled to address the imbalance of outcome. We then train a causal forests using the clinical, demographic and EEG markers for treatment effects estimation. Finally, we estimate the doubly robust scores and employ the efficient policy learning algorithm to learn a optimal policy as a depth-$2$ decision tree.

\noindent $\underline{\textbf{ATE:}}$ We examine the overall efficacy of sertraline treatment on the response rate using estimated ATE from the causal forests. We first estimate $\hat{m}(\cdot)$,  $\hat{e}(\cdot)$ by fitting two separate regression forests and further estimate the conditional average treatment effect $\hat{\tau}(x)$. The tuning parameters are automatically selected via cross validation by minimizing R-loss. We then compute the doubly robust scores $\hat{\Gamma}_i$ using out-of-bag predictions, where:
$$\hat{\Gamma}_i = \hat{\tau}^{(-i)}(X_i) + \frac{W_i-\hat{e}^{(-i)}(X_i)}{\hat{e}^{(-i)}(X_i)[1-\hat{e}^{(-i)}(X_i)]} [Y_i - \hat{m}^{(-i)}(X_i) - [W_i - \hat{e}^{(-i)}(X_i)]\hat{\tau}^{(-i)}].$$ The average treatment effect is estimated using the average of the doubly robust scores: $\hat\tau = \frac{1}{n}\sum^{n}_{i=1} \hat{\Gamma}_i$ and the standard error estimate $\hat{\sigma}^2$ is computed as: $\hat{\sigma}^2 = \frac{1}{n}\sum^{n}_{i=1}(\hat{\Gamma}_i - \hat\tau)^2$. 

The point estimate and $95\%$ confidence interval of the estimated average treatment effect is: $\mathbf{\hat\tau} = \textbf{17.4\% ([2.6\%, 32.2\%])}(\text{p} < \text{0.05})$. And thus the sertraline treatment demonstrates overall efficacy and we could interpret that on average, treating MDD patients with sertraline improves response rate by $17.4\%\ (95\% \text{ CI}:[2.6\%, 32.2\%])$. As a contrast, the proportion of responders in EMBARC is $48.5\%$ and $37.7\%$ in the treatment and placebo arm, respectively. We also note that the ATE estimated by the empirical proportion of response in each group fails to reach statistical significance ($p=0.08$). Since propensity scores are guaranteed to be consistent in a randomized trial, our analysis is free of the risk of model misspecification by the double robustness property. Therefore, this analysis demonstrates that by using double robust scores and adjusting for covariates, we improve the efficiency and reach statistical significance. 

\noindent $\underline{\textbf{HTE:}}$ After estimating the conditional treatment effect $\hat\tau(x)$ via causal forests, we examine its distribution using out-of-bag prediction in the training set and prediction results in the testing set to detect potential evidence of heterogeneity. From the direct observation of the out-of-bag predictions in the training set and predictions in the testing set as shown in \hyperref[figure A4]{Figure A4}, we notice some heterogeneity of the treatment effects.    

In addition to visualization tools, we further adopt a test for heterogeneity motivated by the ``best linear predictor" method of Chernozhukov et al. (2018b\cite{chernozhukov2022generic}. 
Specifically, we seek to estimate the best linear predictor of CATE using out-of-bag causal forests estimates $\hat{\tau}^{-i}(X_i)$ using the following linear model: $$Y_i - \hat{m}^{-i}(X_i) = \alpha\bar{\tau}(W_i - \hat{e}^{-i}(X_i)) + \beta(\hat{\tau}^{-i}(X_i) - \bar{\tau})(W_i - \hat{e}^{-i}(X_i)) + \epsilon_i, $$
where $\bar{\tau}:=\frac{1}{n}\sum^n_{i=1}\hat{\tau}^{-i}(X_i)$. Here $\alpha$ can be used to examine the average prediction by causal forests while $\beta$ can be interpreted as a measure of the quality of the estimates of treatment heterogeneity and we can thus use the p-value associated with the estimate to test for the heterogeneity. If $\alpha = 1$, then the average prediction produced by casual forests is accurate. If $\beta$ is positive and significant, we have evidence of an association between out-of-bag causal forests estimates $\hat{\tau}^{-i}(X_i)$ and the true CATE $\tau(X_i)$. The test results are summarized in \hyperref[table A1]{table A1}. The estimate of $\alpha$ is close to $1$ and the estimates of $\beta$ is positive and significant, we can thus conclude that the average causal forests CATE estimates are correct and that casual forests succeeded in finding heterogeneity in the treatment effect. 

\noindent $\underline{\textbf{Variable Importance:}}$ To further examine the heterogeneity in the treatment effect, we utilize the variable importance measure produced by causal forests as an indication of how often an variable was used in tree splits. The definition of the variable importance in causal forests is as follows: 
$$\text{Importance}_i = \frac{\sum\limits_{l=1}^{\text{max.depth}}\left(\
\frac{\sum_{b=1}^{B}\text{number of splitting in layer }l\text{ for  }x_i\text{ in tree }b}
{\sum\limits_{b=1}^{B}\text{total of splitting in layer }l \text{ in tree }b}
\right)\cdot l^{-2}}{\sum\limits_{l=1}^{\text{max.depth}} l^{-2}}.$$

From the variable importance measures, we can identify features that are important in maximizing the difference in treatment effect between child nodes. As a result, there are many EEG features selected amongst the most important features, indicating the moderating power of those features on the treatment effect. The top three most important features are: (1) fc2.close.theta: eyes close theta band relative power in frontal/central region; (2) c1.open.theta: eyes open theta band relative power in central region; (3) cp5.open.theta: eyes open theta band relative power in central/parietal region. The top ten most important features are listed in \hyperref[figure A5]{Figure A5}.

\begin{figure}
\label{figure 4}
\centering
\includegraphics[width=\textwidth]{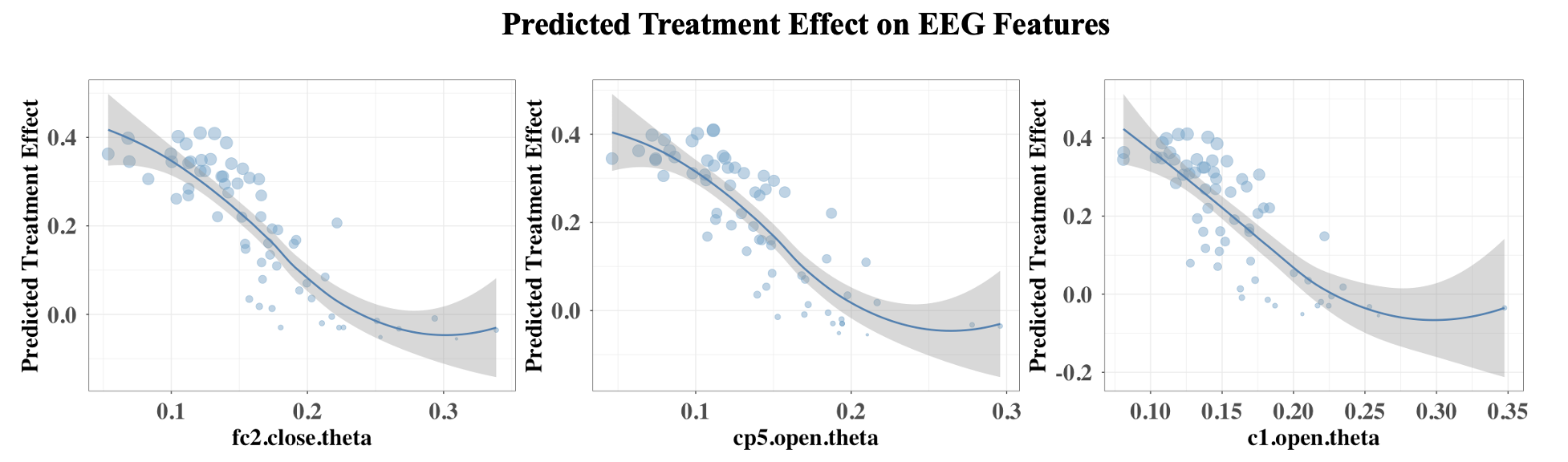}
\caption{Local polynomial regression of predicted treatment effect on important features in the test set with 95\% confidence interval.}
\end{figure}

In addition, we visualize the relationship between these important features and predicted treatment effects in the testing set via local polynomial regression in \hyperref[figure 4]{Figure 4}. We observe that as the eyes close theta band relative power in frontal/central regions increases, the predicted treatment effects decreases. Similarly, as the eyes open theta band relative power in central/occipital region increases, the predicted treatment effects decreases. These results provide interpretation of the dependence between HTE and EEG biomarkers which are useful to guide clinical decisions. 

\begin{figure}
\label{figure 5}
\centering
\includegraphics[width=3.8in]{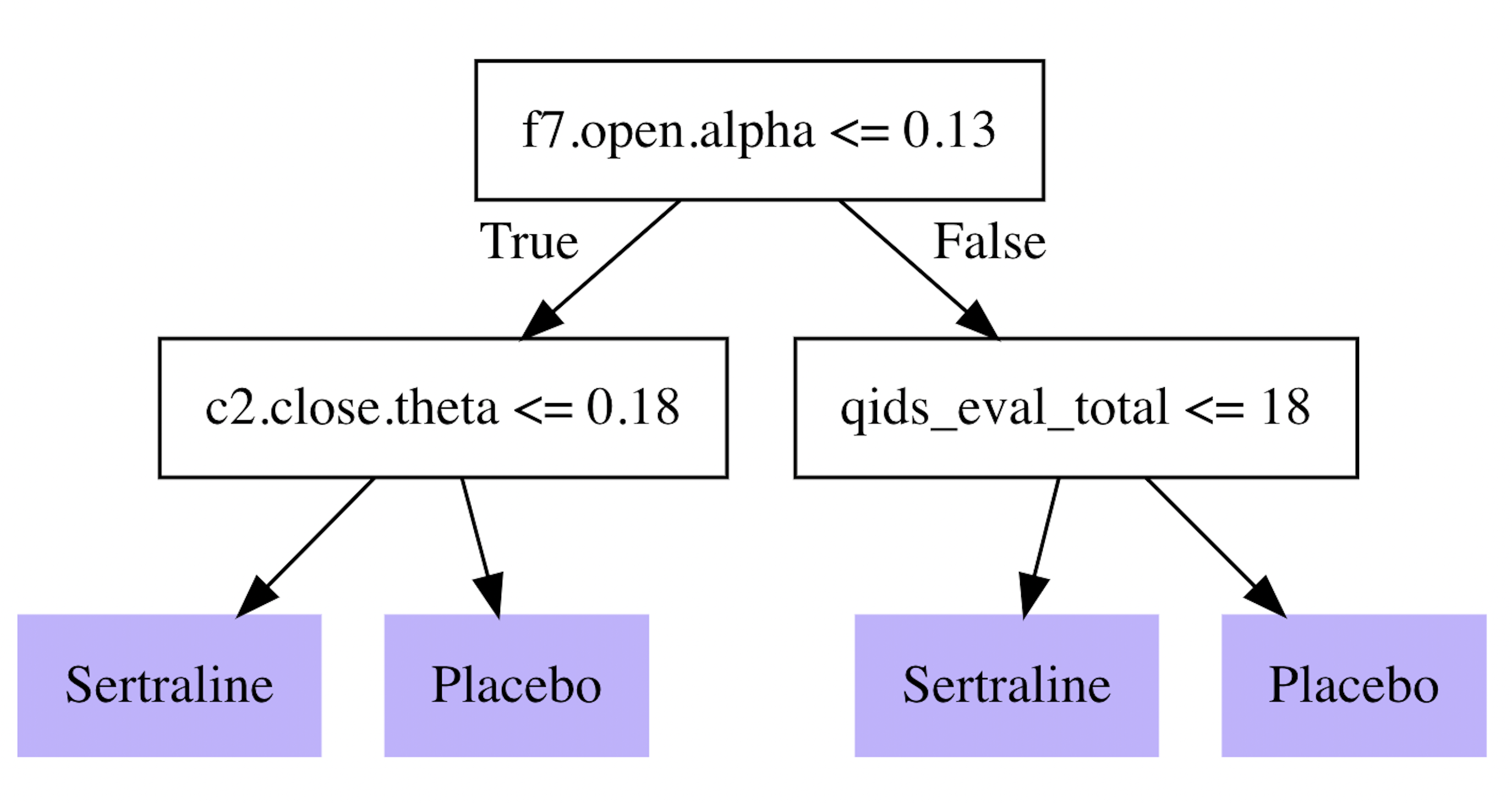}
\caption{Optimal depth-2 policy tree learned by optimizing the augmented inverse propensity weighting loss function.}
\end{figure}

\noindent $\underline{\textbf{Fitted Treatment Assignment Policy:}}$ After obtaining evidence of heterogeneous treatment effect, we propose a treatment assignment policy that selects the best treatment for an individual. We implement algorithms by Athey and Wager (2021)\cite{athey2021policy} via R package $\textbf{policytree}$. The estimating procedure involves three steps: 
\begin{enumerate}
    \item Estimate $\hat{m}(\cdot)$,  $\hat{e}(\cdot)$ by fitting two separate regression forests and further estimate the conditional average treatment effect $\hat{\tau}(x)$ using causal forests. 
    \item Compute the doubly robust scores: $$\hat{\Gamma}_i = \hat{\tau}^{(-i)}(X_i) + \frac{W_i-\hat{e}^{(-i)}(X_i)}{\hat{e}^{(-i)}(X_i)[1-\hat{e}^{(-i)}(X_i)]} [Y_i - \hat{m}^{(-i)}(X_i) - [W_i - \hat{e}^{(-i)}(X_i)]\hat{\tau}^{(-i)}].$$
    \item Select $\hat{\pi} \in \text{argmax}\left\{\sum^{n}_{i=1}(2\pi(X_i)-1)\hat{\Gamma}_i: \pi \in \Pi \right\}$.
\end{enumerate}
We restrict our class of $\Pi$ to be depth-$2$ decision trees for simplicity and interpretability. As indicated by the resulting decision tree \hyperref[figure 5]{Figure 5}, we can implement a treatment assignment policy utilizing EEG features including c2.close.theta (eyes close theta band relative power in central region) and f7.open.alpha (eyes open alpha band relative power in frontal region). 

\begin{figure}
\label{figure 6}
\centering
\includegraphics[width=5in]{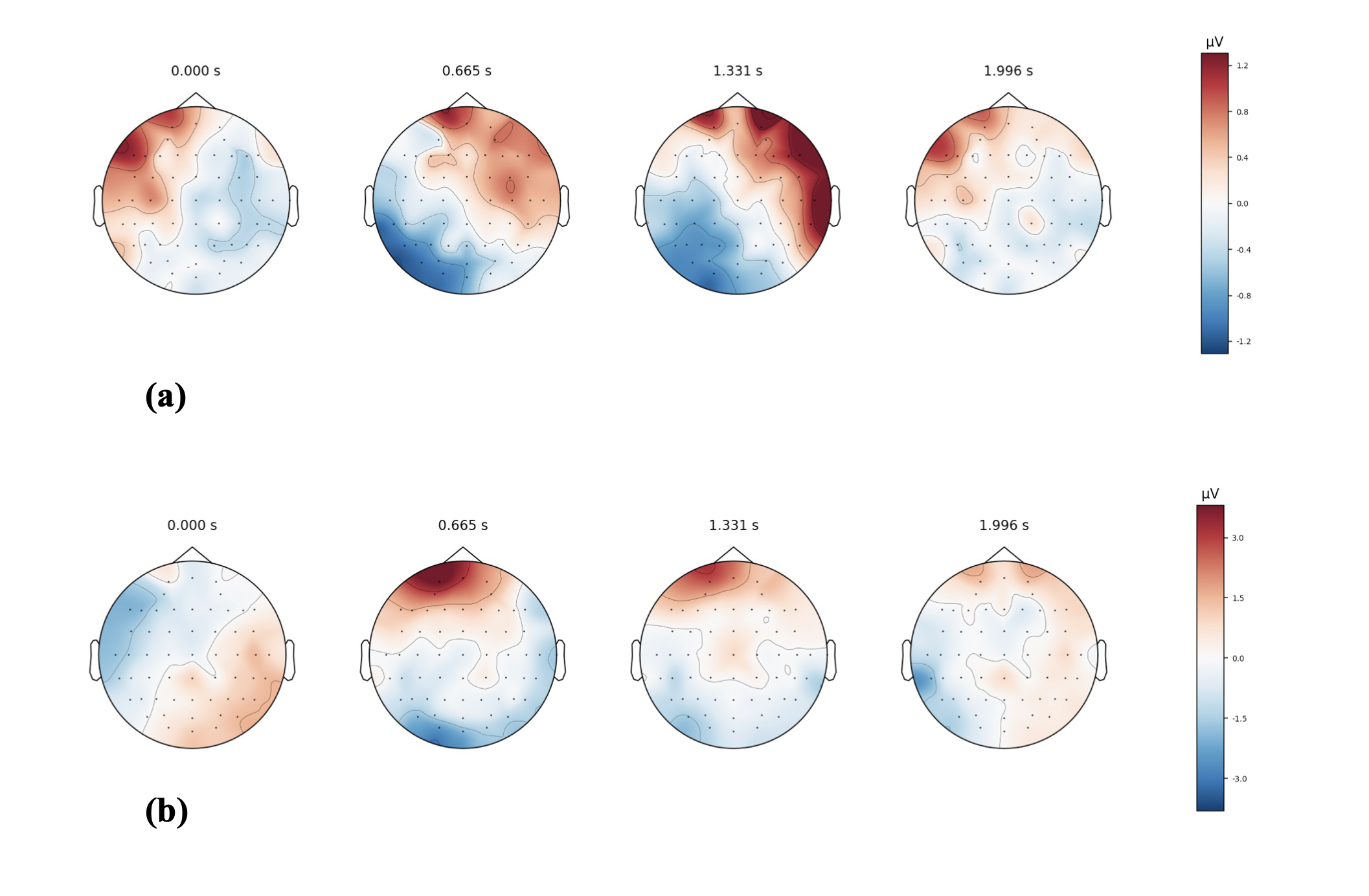}
\caption{(a). Topography for subject whose optimal treatment is Sertraline. (b). Topography for subjects whose optimal treatment is Placebo.  }
\end{figure}

In addition to the treatment assignment policy, we can also visualize the scalp topographies of subjects assigned to setraline and placebo as shown in \hyperref[figure 6]{Figure 6}. We average EEG signal over the epochs in the first eyes open segment, and visualize the power of the averaged signal of subject assigned to sertraline and placebo. From the scalp topographies, we see a drastic difference in the EEG power distribution on the scalp, demonstrating the moderating effect of EEG features for treatment assignment. 

Finally, we compare the performance of our employed algorithm with Q-learning implemented with Lasso and O-learning with binomial deviance loss in terms of the value function $V(\pi) = \mathbb{E}[Y_i(\pi(X_i))]$ via a three-fold cross validation. As summarized in the \hyperref[table A2]{table A2}, causal tree performed better than Q-learning and O-learning.

\noindent $\underline{\textbf{Comparison of Raw and Processed Features:}}$ In the last section, we present evidence that features extracted from our EEG preprocessing pipeline possess stronger signal compared with raw features. We train causal forests on the processed features set and raw features set separately, and compare their prediction performance on CATE via mean squared error $\mathbb{E}[(Y_i^* - \hat{\tau}(X_i))^2]$ computed using a transformed outcome, $Y_i^{*} = (Y_i - W_i)/(p \cdot (1 - p)),$ as discussed in Tian et al. (2014)\cite{tian2014simple}. As shown in \hyperref[table A3]{table A3}, we obtained better prediction using the processed features, demonstrating the utility of our EEG cleaning pipeline. 

\section{Discussion}
\label{Section 5}
In this work, we aim to design an integrated pipeline for treatment effect estimation and policy learning that improves the response rate of MDD by utilizing the resting state, pre-treatment EEG measures.  Through an application of our pipeline to a large multi-site, double-blind randomized controlled clinical trial (EMBARC \citep{trivedi2016establishing}), we first tackle the challenges in EEG preprocessing by incorporating two steps that are not seen in the previous studies: automatic bad channel identification, interpolation and a data-driven bad epoch rejection. We achieve an automatic EEG processing and feature extraction pipeline that demonstrates its utility through contrast in correlation and reduced model MSE in raw and processed features. 
 
 To handle the high-dimensionality of EEG biomarkers, we employ casual forests for a powerful estimation of the CATE and present the evidence of EEG features modifying treatment through important features in splitting as well as a significant $p$-value in the best linear predictor test. In addition, the causal forests also provides an efficient doubly robust estimation of the average treatment effect and we detect a significant ATE for response rate at the end of study, demonstrating the efficacy of setraline treatment. Finally, we further utilize the doubly robust scores and employ an efficient policy learning algorithm that estimates an optimal policy as a depth-2 decision tree. We demonstrate the superiority of this algorithm by comparing the value function with Q-learning implemented with Lasso and O-learning with binomial deviance loss in a three-fold cross validation as well as in several synthesized simulation studies.  

A limitation of this work is studying a single neuroimaging modality. Functional MRI measures were not included due to a large number of subjects with missing data. It is of interest to consider multi-modal fusion and best ways to handle missing modality in the context of treatment policy learning. In addition, future studies may examine multi-stage clinical trials and validation in an independent cohort. 

In conclusion, we present evidence that non-invasive EEG measures of relative theta and alpha relative band power could aid detecting heterogeneous treatment effects and learning an optimal treatment assignment policy for depression. Particularly, we show that our automatic EEG preprocessing and feature extraction procedure yields features that possess stronger signal. The EEG processing and feature extraction procedure is unified with heterogeneous treatment effects estimation and efficient policy learning in our integrated pipeline. This streamlined estimation procedure may assist clinicians to assign treatments using a simple decision tree depending on EEG features.

\section*{Acknowledgments}       
This research is supported by U.S. NIH grants MH123487, NS073671, and GM124104.

\bibliography{myref}

\appendix

\section{TABLES AND ADDITIONAL FIGURES\label{app1}}

We present tables and additional figures that are referenced in this section. 

\begin{figure}[ht]
\label{figure A1}
\centering
\includegraphics[width=6in]{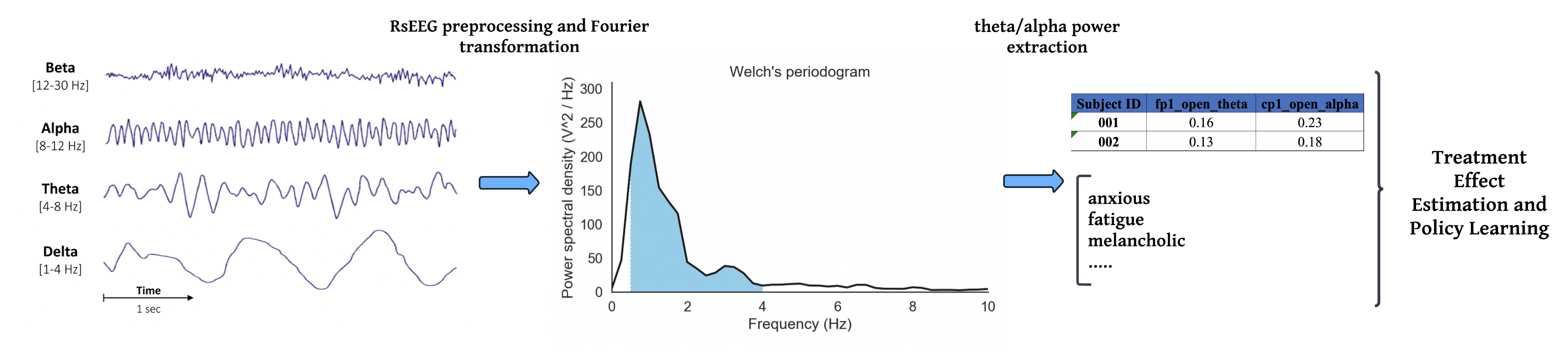}
\caption{Analysis Aim to Leverage rsEEG and Clinical Markers
for Treatment Effect Estimation and Policy Learning}
\end{figure}

\begin{figure}[ht]
\label{figure A2}
\centering
\includegraphics[width=5in]{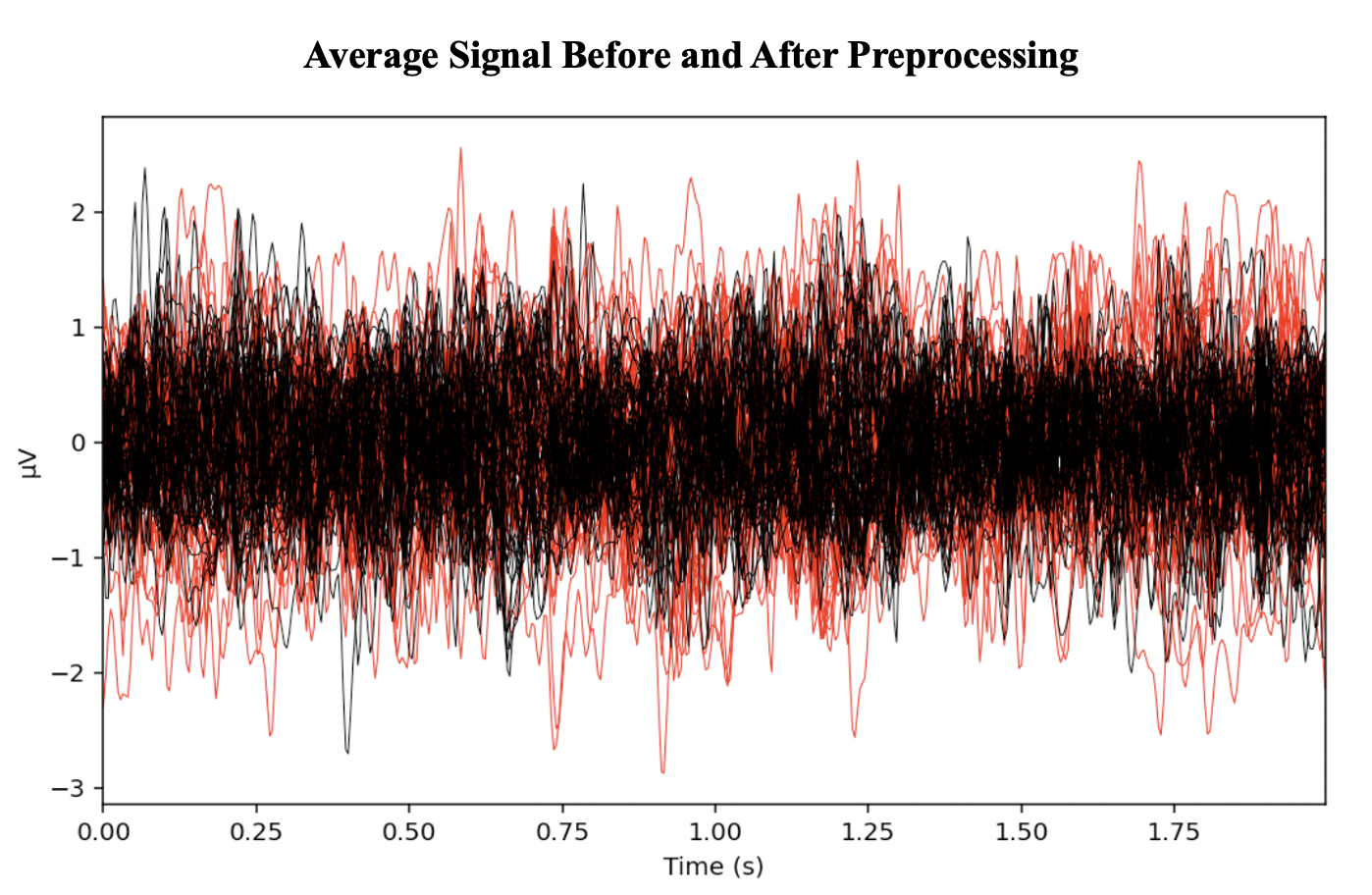}
\caption{Average signal before and after Preprocessing. Red lines denote average signal of raw data and black lines denote the average signal for processed data.}
\end{figure}

\begin{figure}[ht]
\label{figure A3}
\centering
\includegraphics[width=5in]{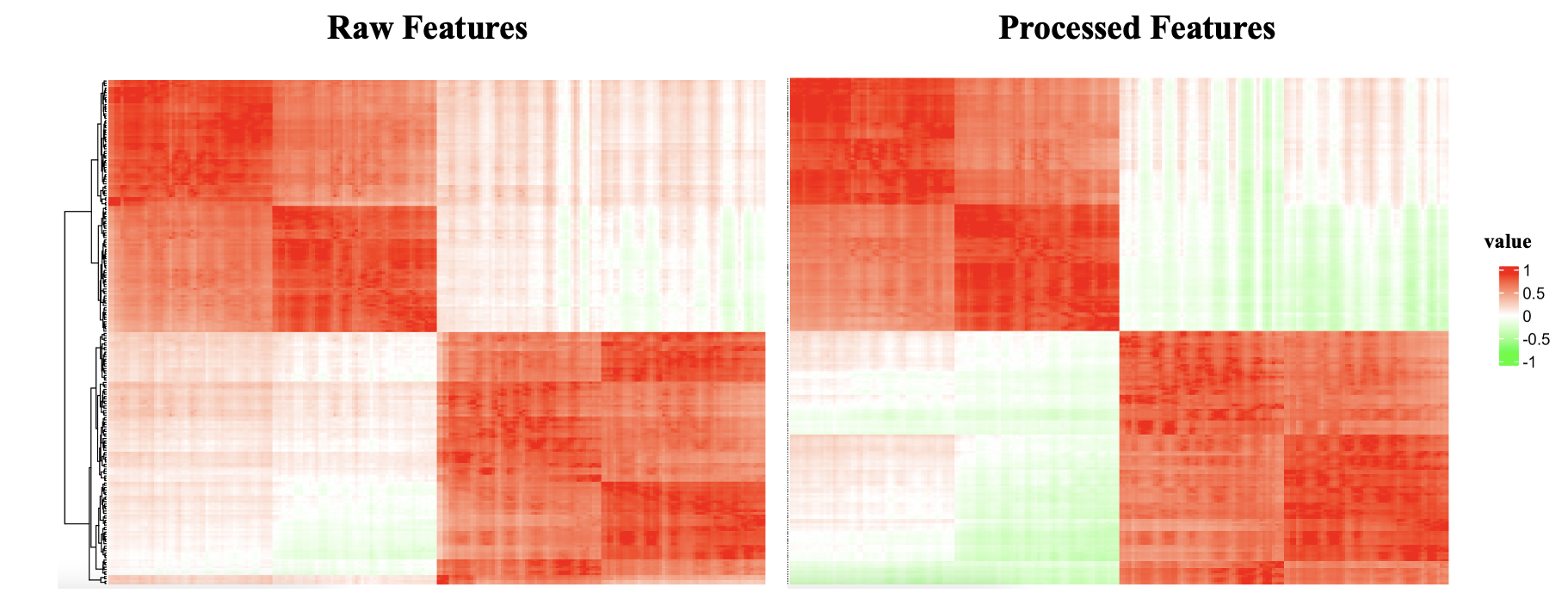}
\caption{Left-hand side plot visualizes correlation matrix for features extracted from raw data set and right-hand side plot summarizes correlation matrix for features extracted from processed data set.}
\end{figure}

\begin{figure}
\label{figure A4}
\centering
\includegraphics[width=6in]{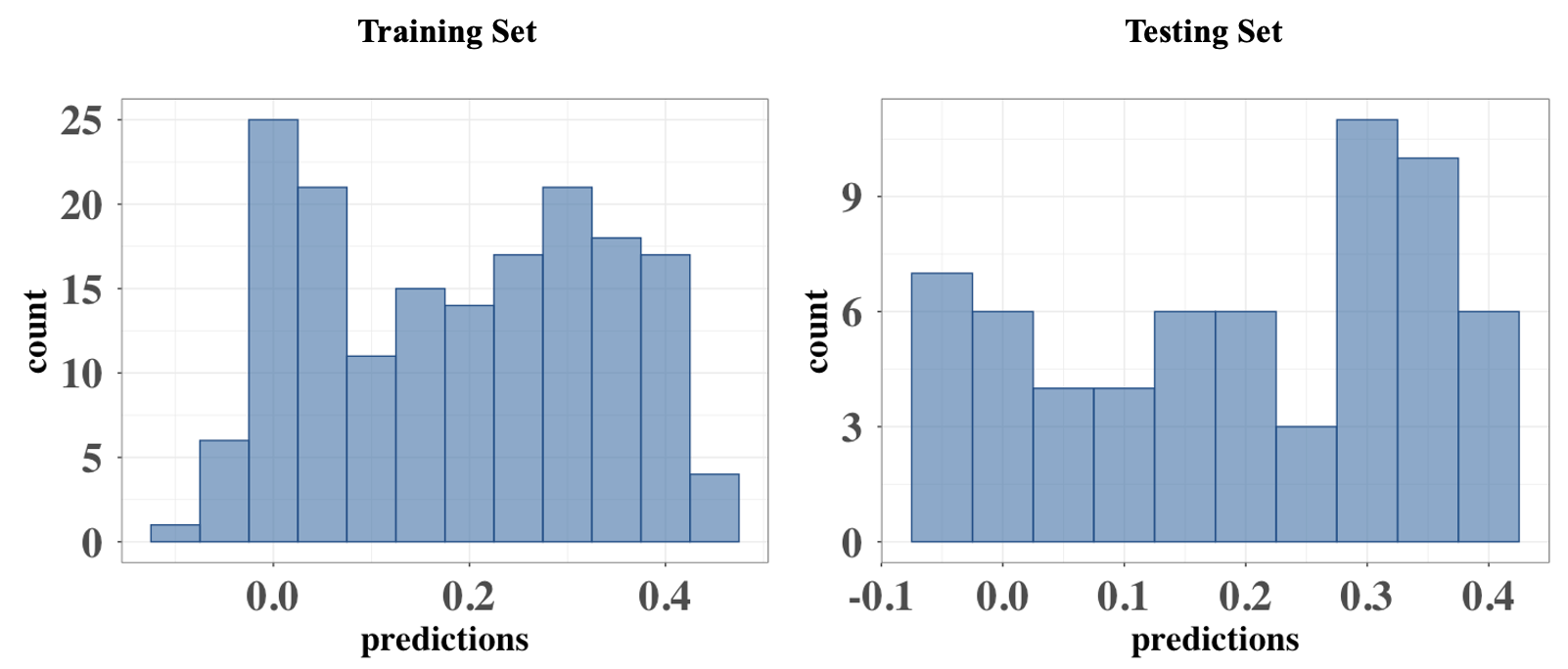}
\caption{Histogram of Prediction of Treatment Effects}
\end{figure}

\begin{figure}
\label{figure A5}
\centering
\includegraphics[width=4.5in]{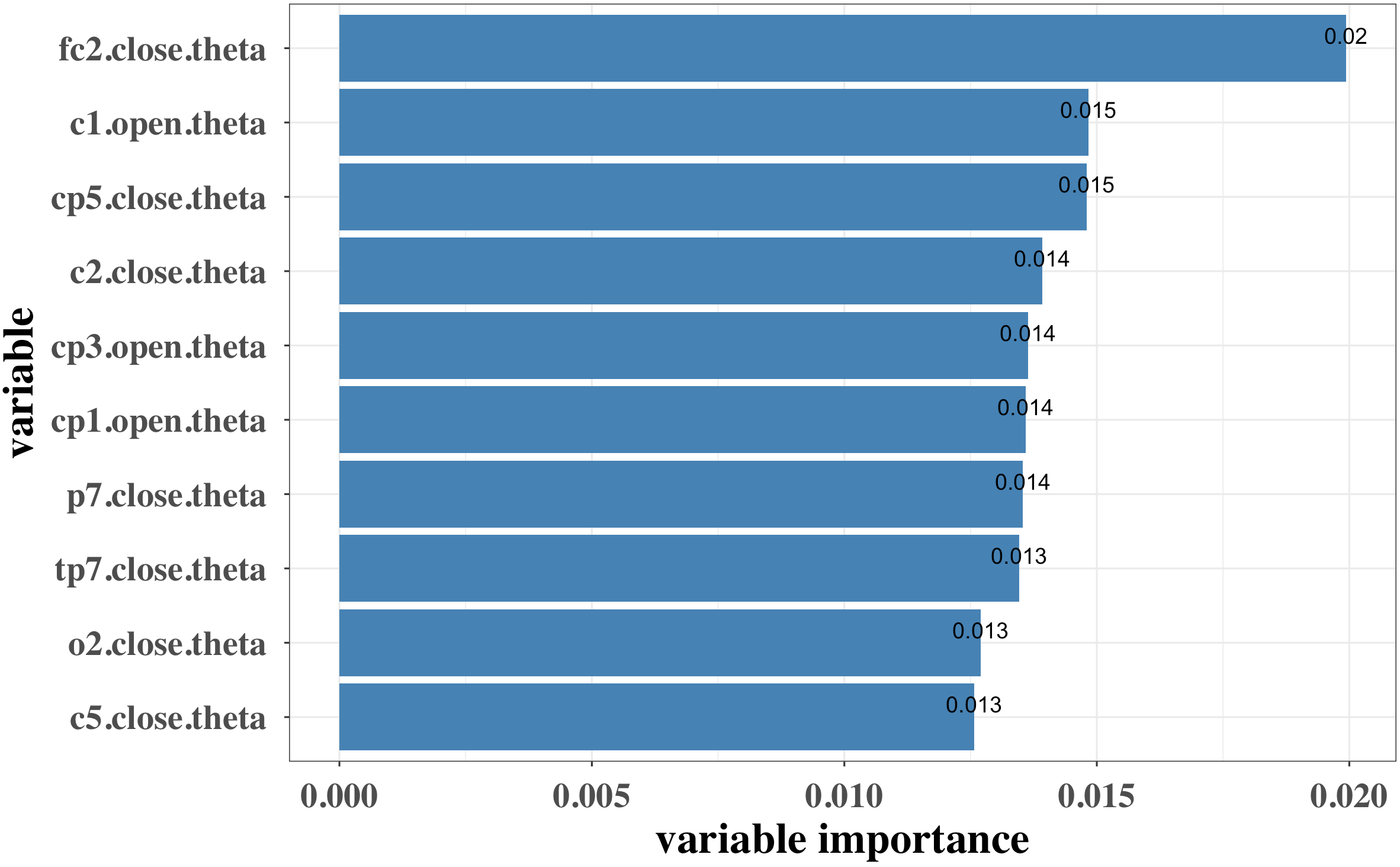}
\caption{Variable importance measure of the top ten most important features}
\end{figure}

\begin{center}
\label{table A1}
\begin{table}[ht]
\caption{Best Linear Predictor Test Results}
\centering
\begin{tabular*}{500pt}{@{\extracolsep\fill}lcccc@{\extracolsep\fill}}
\toprule
\textbf{} & \textbf{Estimates}  & \textbf{Standard Error}  & \textbf{t value}  & \textbf{Pr(\(>t\))} \\
\midrule
$\alpha$ & $0.912$ & $0.398$ & $2.292$ & $0.0116$ \\
$\beta$ & $1.418$ & $0.518$ & $2.736$ & $0.00345$ \\
\bottomrule
\end{tabular*}
\end{table}
\end{center}

\begin{center}
\label{table A2}
\begin{table}[ht]
\caption{Mean and standard error of value function estimate of depth-2 policy tree, O-learning and Q-learning estimated from a three-fold cross validation.}
\centering
\begin{tabular*}{500pt}{@{\extracolsep\fill}lccc@{\extracolsep\fill}}
\toprule
\textbf{} & \textbf{Mean value}  & \textbf{Standard Error}   \\
\midrule
Policy Tree & 0.475 & 0.0604 \\     
 O-learning & 0.459 & 0.0503 \\
 Q-learning & 0.439 & 0.0230 \\ 
\bottomrule
\end{tabular*}
\end{table}
\end{center}

\begin{center}
\label{table A3}
\begin{table}[t]
\caption{Mean squared error and standard error of transformed outcome of causal forests estimated using processed and raw features.}
\centering
\begin{tabular*}{500pt}{@{\extracolsep\fill}lccc@{\extracolsep\fill}}
\toprule
\textbf{} & \textbf{Mean value}  & \textbf{Standard Error}   \\
\midrule
 Processed features & 1.84 & 2.09 \\ 
 Raw features & 1.94 & 2.48  \\
\bottomrule
\end{tabular*}
\end{table}
\end{center}

\end{document}